\documentclass[%
 twocolumn,
 groupedaddress
 amsmath,amssymb,
 aps,
 prl
]{revtex4}

\usepackage{graphicx}
\usepackage{dcolumn}
\usepackage{bm}
\usepackage{here}
\usepackage{color}
\usepackage{ulem}
\usepackage{textcomp}
\usepackage{amsmath}
\usepackage{nccmath}

\begin{document}

\preprint{APS/123-QED}

\title{Direct coupling of ferromagnetic moment and ferroelectric polarization in BiFeO$_{3}$}
\author{Shiro Kawachi,$^1$ Shin Miyahara,$^2$ Toshimitsu Ito,$^3$ Atsushi Miyake,$^4$ Nobuo Furukawa,$^5$ Jun-ichi Yamaura,$^{1}$ and Masashi Tokunaga$^4$}
\affiliation{\vspace{4pt}$^1$Materials Research Center for Element Strategy, Tokyo Institute of Technology, Yokohama, Kanagawa 226-8503, Japan.
\\$^2${Department of Applied Physics, Fukuoka University, Jonan-ku, Fukuoka 814-0180, Japan.}
\\$^3$National Institute of Advanced Industrial Science and Technology (AIST), Tsukuba, Ibaraki 305-8562, Japan.
\\$^4$The Institute for Solid State Physics, The University of Tokyo, Kashiwa, Chiba 277-8581, Japan.
\\$^5$Department of Physics and Mathematics, Aoyama Gakuin University, Sagamihara, Kanagawa 229-8558, Japan.
}
\begin{abstract}
The spin-driven component of electric polarization in a single crystal of multiferroic BiFeO$_{3}$ was experimentally investigated in pulsed high magnetic fields up to 41 T.
Sequential measurements of electric polarization for various magnetic field directions provide clear evidence of electric polarization normal to the hexagonal $c$ axis (${\bm P}_{\rm t}$) in not only the cycloidal phase, but also the field-induced canted antiferromagnetic phase. 
The direction of ${\bm P}_{\rm t}$ is directly coupled with the ferromagnetic moment in the canted antiferromagnetic phase, and thus controlled by changing the direction of the applied magnetic field. This magnetoelectric coupling is reasonably reproduced by the metal--ligand hybridization model.
\end{abstract}

\maketitle

\par The crossed coupling of magnetism and ferroelectricity in multiferroic materials has unified two major research areas in condensed matter physics that were traditionally distinguished. 
In addition to being of interest from a point of view of basic science, magnetoelectric coupling is expected to allow the development of innovative devices, such as electrically controllable magnetic memory and multi-bit memory \cite{Bibes2008, Scott2007}. While most multiferroic materials are active only at low temperatures, BiFeO$_3$, which exhibits remarkable multiferroic properties at room temperature, has attracted much interest \cite{Catalan2009}.
\par BiFeO$_{3}$ crystallizes in the non-centrosymmetric $R3c$ space group within a trigonal crystal system at temperatures below $\sim 1100$ K \cite{Kubel1990}.
Atomic displacement from the perovskite-type structure results in spontaneous giant electric polarization ${\bm P}_{\rm s}$ along the hexagonal $c$ axis, which corresponds to $\langle111\rangle_{\rm c}$ in cubic perovskites [the $+Z$ direction in Fig. 1(a)] \cite{Catalan2009, Shvartsman2007, Lebeugle2007}.
The $S = 5/2$ spins of ${\rm Fe}^{3+}$ ions form a cycloidal magnetic order with the $XY$-component of the magnetic propagation vector ${\bm q}$ pointing in the hexagonal direction $\langle100\rangle_{\rm h}$ below temperatures of $\sim 640$ K \cite{Sosnovska1982}.
Three-fold rotational symmetry allows the coexistence of three equivalent magnetic domains \cite{Johnson2013}.
The direction of $\bm{q}$ can be controlled by flopping ${\bm P}_{\rm s}$ with an electric field \cite{Lebeugle2008, Lee2008PRB}. 
This flop of ${\bm P}_{\rm s}$ is, however, accompanied by crystal deformation and hence often degrades the crystal, which is unfavorable for application. 
\begin{flushleft}
\begin{figure}[t]
 \centering
 \includegraphics[width=8.6cm]{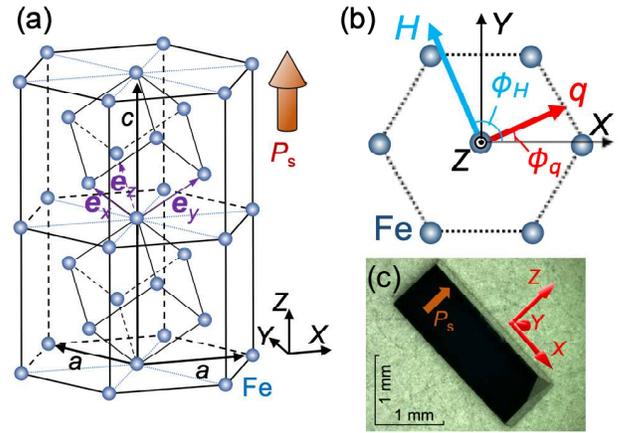} 
 \caption{(a) Illustration of the arrangement of Fe ions in BiFeO$_{3}$. ${\bm a}$ and ${\bm c}$ represent unit vectors in the hexagonal cell. ${\bm P}_{\rm s}$ indicates spontaneous ferroelectric polarization below $\sim$1100 K. 
(b) Schematic drawing of the definition of azimuthal angles for the applied field and cycloidal ${\bm q}$ vector, $\phi_{H}$ and $\phi_{q}$ respectively. (c) Photograph of the single crystal used for the measurements. 
}\label{fig:fig1}
\end{figure}
\end{flushleft}
\par In addition to ${\bm P}_{\rm s}$, BiFeO$_{3}$ has parasitic ferroelectric polarization originating from the magnetic order.
In the cycloidal phase, BiFeO$_{3}$ undergoes a magnetic phase transition with the application of a magnetic field of about 20 T \cite{Kadomtseva2004}.
Extrapolation of the magnetization curve in the field-induced phase to a zero magnetic field has a finite intercept, which suggests the occurrence of the canted antiferromagnetic (CAFM) order with a net ferromagnetic moment of $ \sim 0.04$ {\textmu}$_{\rm B}$/Fe on the $XY$ plane \cite{Kadomtseva2004, Ruette2004, Pyatakov2009, Tokunaga2015, Kawachi2017}.
In general, the cycloidal magnetic order is known to be accompanied by spin-driven electric polarization \cite{Katsura2005, Arima2011, Kaplan2011, Pyatakov2015, Miyahara2016}.
Field-induced suppression of this phase may affect the electric polarization.
Recent experiments using single crystals revealed spin-driven electric polarization in two directions longitudinal and transverse to the $Z$ direction \cite{Park2011, Lee2013, Lee2015, Tokunaga2015}.
The transverse electric polarization (${\bm P}_{\rm t}$) in the cycloidal state is explained by the inverse Dzyaloshinskii--Moriya (DM) model \cite{Lee2015, Miyahara2016, Tokunaga2015}, with which the emergence of ${\bm P}_{\rm t}$ normal to the ${\bm q}$-${\bm Z}$ plane in a zero field is predicted. Reorientation of magnetic domains is accompanied by the reorientation of ${\bm P}_{\rm t}$. 
This effect appears as a non-volatile memory effect that can be controlled by a magnetic field or electric field even at room temperature \cite{Tokunaga2015, Kawachi2016}. 
\par In this study, we determined the direction and absolute value of ${\bm P}_{\rm t}$ and its field-angle dependence in a single crystal of BiFeO$_3$ through a series of magnetoelectric measurements by changing the direction of the applied magnetic field. 
The results revealed the doubling of ${\bm P}_{\rm t}$ at the transition from the cycloidal to the CAFM state. 
Semi-quantitative explanation of the results by the metal--ligand hybridization mechanism clarified the microscopic origin for the spontaneous coupling between ferromagnetism and ferroelectricity in this material, and demonstrated that prominent multiferroic coupling could be realized without complicated spin order. 
\par Single crystals of BiFeO$_{3}$ were grown adopting the laser-diode heating floating-zone method \cite{Ito2011}. The obtained crystal almost entirely consisted of a single ferroelectric domain of ${\bm P}_{\rm s}$. 
A crystal used for the measurements was cut into a rectangular parallelepiped shape of $2.34\times 0.49\times 0.85$ mm$^3$ along the crystal axes as shown in Fig. 1(c). 
To determine the crystal orientation with its chirality, we conducted an x-ray diffraction measurement using a curved imaging plate on a RIGAKU VariMax with RAPID diffractometer at the wavelength of Mo-K$\alpha$ radiation.
The crystal chirality was identified by the intensities of Bijvoet-pair-related reflections of $\{6\bar{3}6\}_{\rm h}$ and $\{6\bar{3}\bar{6}\}_{\rm h}$ on a hexagonal setting owing to the loss of inversion symmetry based on atomic coordinates given in the literature \cite{Kubel1990} (See Supplementary Material).
The determined $a$ and $c$ hexagonal axes are chosen as the $+X$ and $+Z$ directions. By this definition, ${\bm P}_{\rm s}$ points in the $+Z$ direction as shown in Fig. 1(a).
Magnetic-field-induced changes in electric polarization were measured by integrating (de)polarization currents \cite{Mitamura2007} induced in pulsed magnetic fields generated using a 55-T magnet for a duration of 8 ms.
These experiments only determined the relative changes in electric polarization ($\Delta P$) with respect to the initial values for a zero field. 
To determine the zero level of electric polarization, we performed a series of experiments while changing the field angle using a sample rotator as discussed in the following.
\begin{flushleft}
\begin{figure}[t]
 \centering
 \includegraphics[width=8.6cm]{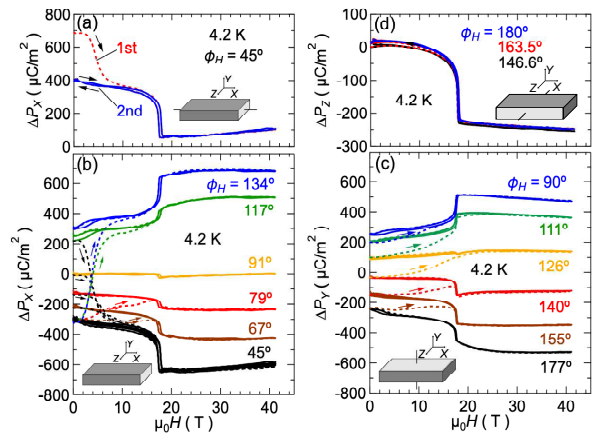} 
 \caption{
 (a) Magnetic-field-induced changes in electric polarization along $X$ at $\phi_{H}=45^{\circ}$. Traces in the first and second cycles are respectively shown by dotted and solid lines. 
Magnetic-field-induced changes in electric polarization along (b) $X$, (c) $Y$, and (d) $Z$ directions, measured at various field angles.
Traces are vertically offset so that the initial point of each curve is the end point of the previous curve.
The inset presents a rectangular parallelepiped sample of BiFeO$_{3}$ (dark gray area) with electrodes (light gray area) to visualize the measurement direction.
 }\label{fig:fig2}
\end{figure}
\end{flushleft}
%
\par Typical results are shown in Fig. 2(a). 
Prior to this experiment, we applied a magnetic field ($H$) with a magnitude up to 41 T at an azimuthal angle of $\phi_{H}=91^{\circ}$ ($\sim H\parallel Y$) as defined in Fig. 1(b). 
We then measured $\Delta P$ along the $X$ direction ($\Delta P_{X}$) as a function of the magnetic field at $\phi_{H}=45^{\circ}$. 
In the first field cycle, $\Delta P_{X}$ has a hysteretic trace below 10 T [red dotted line in Fig. 2(a)], which is ascribed to the reorientation of the magnetoelectric domains. 
Here, the change in $P_{X}$ from 0 to 20 T becomes small after reorientation of the magnetoelectric domain. The generalized inverse DM model predicts the emergence of the largest $P_{X}$ in the reoriented state and absence of it in the CAFM state. The observed reduction in $\Delta P_{X}$ is therefore a puzzling phenomenon \cite{Tokunaga2015}.
\par To unambiguously determine the spin-driven components in electric polarization, we sequentially measured $\Delta P_{X}$ while systematically changing $\phi_{H}$.
Representative data of $\Delta P_{X}$, $\Delta P_{Y}$, and $\Delta P_{Z}$ are presented in Figs. 2(b)--(d) as functions of $H$ applied at various $\phi_{H}$. All data presented in the following were recorded at a temperature of 4.2 K to minimize the effect of leakage current.
Throughout this letter, $P$--$H$ curves obtained in the first and second magnetic field sweeps for each $\phi_{H}$ are respectively represented by dotted and solid lines. 
In Figs. 2(b) and 2(c), all traces are offset on the vertical axis so that the starting point of the trace coincides with the endpoint of the previous trace.
The curves of $\Delta P_{X}$ in Fig. 2(b) were measured in the order of $\phi_{H}=91^{\circ}\rightarrow45^{\circ}\rightarrow67^{\circ}\rightarrow45^{\circ}\rightarrow79^{\circ}\rightarrow45^{\circ}\rightarrow117^{\circ}\rightarrow45^{\circ}\rightarrow134^{\circ}$. 
We here determined the zero level of $P_{X}$ by taking the average of the largest and smallest values of $P_{X}$ for a zero field. 
Adopting the present definition of the zero level, we can reasonably reproduce the observed angular dependence using a theoretical model as will be seen later.
\par The data at $\phi_{H}=134^{\circ}$ were obtained after application of a field at $\phi_{H}=45^{\circ}$. Thereby, an irreversible change in $P_{X}$ below 10 T in the first field sweep is due to the $90^{\circ}$ rotation of the ${\bm q}$ vector, which appears as a sign change in $P_{X}$ as shown by the blue dotted line. A steep change in $P_{X}$ at $\sim 18$ T corresponds to the transition from the cycloidal phase to the CAFM phase. The largest value of $P_{X}$ in the cycloidal state is thus determined as $\sim 300$ {\textmu}C/m$^{2}$ at a zero field. Meanwhile, $P_{X}$ in the CAFM phase is almost twice this value.
\par Moreover, we also carried out for $P_{Y}$ on the same crystal in the order of $\phi_{H}=177^{\circ}\rightarrow155^{\circ}\rightarrow140^{\circ}\rightarrow126^{\circ}\rightarrow111^{\circ}\rightarrow90^{\circ}$. 
The results are plotted in Fig. 2(c) adopting a vertical offset as in the previous case. 
Here, the zero level of $P_{Y}$ is defined as the average value of $P_{Y}$ for the zero field after the application of fields along $\phi_{H}=90^{\circ}$ and $177^{\circ}$. 
We again observe larger $P_{Y}$ in the CAFM phase for fields stronger than 18 T. 
\par Meanwhile, $\Delta P_{Z}$ has negligible angular dependence as shown in Fig. 2(d). 
In this case, we can hardly determine the zero level for $P_{Z}$. $P_{Z}$ steeply decreases by $\sim 240$ {\textmu}C/m$^{2}$; i.e., total $P_{Z}$ slightly reduces at the cycloid-to-CAFM transition around 18 T independently of the azimuthal angle of the applied field.
%
\begin{flushleft}
\begin{figure}[t]
 \centering
 \includegraphics[width=8.6cm]{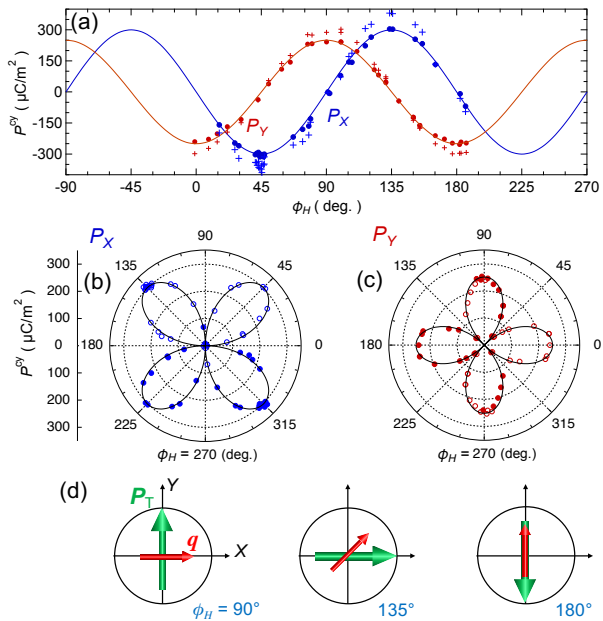} 
 \caption{
(a) Angular dependence of the electric polarization along $X$ and $Y$ directions at \textmu$_{0}H=0$ and 10 T in the cycloidal phase shown by solid circles and crosses, respectively. 
The solid line indicates the fitting function of eq. (3) and/or (4) for solid circles at 0 T. 
 Polar plots presenting the dependencies of (b) $P_{X}$ and (c) $P_{Y}$ on the magnetic field angle after removal of the magnetic field in the cycloidal phase.
 Open symbols are projected points of the experimental data at $\phi_{H}$ (solid symbols) to $\phi_{H}+\pi$.
 (d) Schematics of relative directions between ${\bm P}_{\rm t}$ and the $XY$ component of the ${\bm q}$ vector in the cycloidal phase.
 }\label{fig:fig3}
\end{figure}
\end{flushleft}
\par Figure 3(a) presents $P_{X}$ and $P_{Y}$ as functions of $\phi_{H}$ at 0 T in the cycloidal phase ($P^{\rm cy}$). 
Here, the data were extracted in the reoriented domain state after the first field cycle at each $\phi_{H}$. 
According to Bordacs $et$ $al$., the direction of the ${\bm q}$ vector for a zero field may slightly deviate from the direction perpendicular to the applied field owing to small anisotropy within the $XY$ plane \cite{Bordacs2018, Fishman2018}. 
This deviation, however, plays only a minor role because $P_{X}$ and $P_{Y}$ for a field strength of 10 T, which is higher than the strength of pinning threshold fields of $\sim 5$ T \cite{Bordacs2018}, follow almost the same functional forms as shown by crosses in Fig. 3(a). 
Figures 3(b) and 3(c) exhibit polar plots of the $P_{X}$ and $P_{Y}$ shown in Fig. 3(a). 
The solid circles represent the experimental data, while the open ones are projection of the data at $\phi_H$ to $\phi_H + \pi$. 
For some field angles, we confirmed that the magnetoelectric effect appears symmetric with respect to the sign change in the applied field as shown in Supplementary Material.
\par Continuous changes in $P_{X}$ and $P_{Y}$ indicate that $\phi_{q}$ smoothly follows changes in $\phi_{H}$ in the cycloidal state.
As for the slight difference between the amplitudes of $P_{X}$ and $P_{Y}$, we cannot rule out possible error in the estimation of the effective surface area of the sample. 
$P_{X}$ and $P_{Y}$ respectively trace simple functions of $-P_{\rm t}\sin{2\phi_{H}}$ and $-P_{\rm t}\cos{2\phi_{H}}$. 
This means that ${\bm P}_{\rm t}$ rotates in the $XY$ plane while the magnitude is almost unchanged. 
\par Figure 3(d) illustrates the relationship between ${\bm P}_{\rm t}$ and ${\bm q}$ on the $XY$ plane determined by the experimental data at several magnetic field angles in the cycloidal phase; ${\bm q}\perp {\bm P_{\rm t}}$ at $\phi_{H}=(1/2 + 2n/3)\pi$ and ${\bm q}\parallel {\bm P_{\rm t}}$ at $\phi_{H}=2{\pi}n/3$ with integer $n$. 
${\bm P_{\rm t}}$ rotates in the opposite direction to the magnetic field. 
\begin{flushleft}
\begin{figure}[t]
 \centering
 \includegraphics[width=8.6cm]{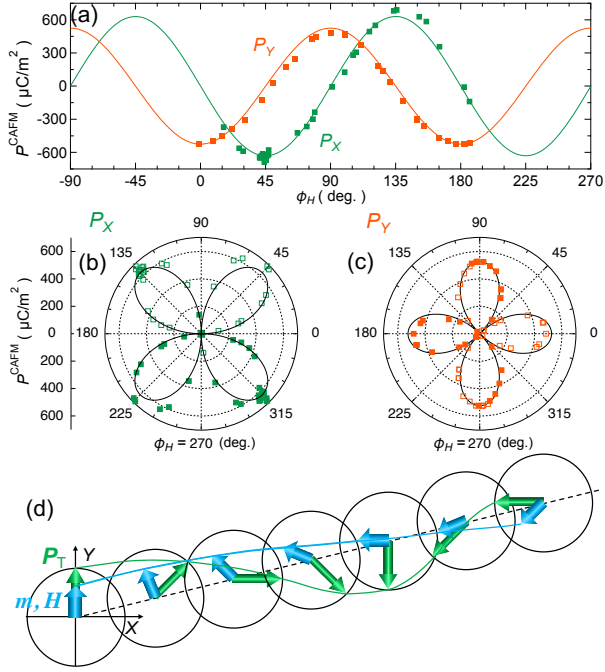} 
 \caption{ (a) Field-angle dependence of the electric polarization along $X$ (orange) and $Y$ (green) directions at \textmu$_{0}H=40$ T in the CAFM phase. 
The solid line shows the fitting functions of $P_{X}$ and $P_{Y}$ obtained using eq. (3). 
Polar plots represent the magnetic-field-angle dependencies of (b) $P_{X}$ and (c) $P_{Y}$ at 40 T in the CAFM phase. 
Experimental data and a theoretical curve are respectively shown by symbols and a solid line. 
 The open symbols are a projection of experimental data at $\phi_{H}$ (solid symbols) to $\phi_{H}+\pi$. (d) Schematic relation between the ferromagnetic moment (blue arrow) and ${\bm P}_{\rm t}$ (green arrow) in the CAFM phase. 
 }\label{fig:fig4}
\end{figure}
\end{flushleft}
\par Figures 4(a)--4(c) show the $\phi_{H}$ dependencies of $P_{X}$ and $P_{Y}$ at 40 T in the CAFM phase. $P_{X}$ and $P_{Y}$ have the same functional formula as in the cycloidal phase. 
The relative orientation determined experimentally between the ferromagnetic component in the CAFM state (${\bm m}$) and ${\bm P}_{\rm t}$ is schematically shown in Fig. 4(d). 
${\bm m}$ follows a curve parallel to ${\bm H}$ owing to the Zeeman energy \cite{Tokunaga2015, Kawachi2017}, while ${\bm P}_{\rm t}$ rotates in the opposite direction to the rotation of ${\bm H}$ by a double angle. This result means that if we rotate ${\bm m}$ at a frequency of $\omega$, we will have rotation of ${\bm P}_{\rm t}$ at the frequency of $2\omega$. 
\par Let us discuss the origin for ${\bm P}_{\rm t}$ coupled with the magnetic order. The multiferroic behavior is usually reproduced using spin-dependent electric polarization:
\begin{eqnarray}
  {p}^\alpha_i & = & \sum_{\beta \gamma} 
  \Lambda^{\alpha}_{\beta \gamma} S_i^\beta  S_{i}^\gamma \\
  {p}^\alpha_{i\, j} & = &
  \Pi_{ij}^\alpha {\bm S}_i \cdot {\bm S}_{j} 
  + \sum_\beta d_{ij}^{\alpha \beta} \left({\bm S}_i \times {\bm S}_{j} \right)^\beta,
\end{eqnarray}
where $\alpha, \beta, \gamma = X, Y, Z$ and $\Lambda, \Pi_{ij}, d_{ij}$ are coupling constants \cite{Moriya1968}. 
${p}_i$ is a single-spin term that is induced by metal--ligand hybridization (ML) and ${p}_{ij}$ is a spin pair term that includes the electric polarization induced by symmetric exchange striction (ES) and antisymmetric inverse DM interaction.
We consider ${\bm p}_{i}$ at Fe ion sites and 
${\bm p}_{ij}$ for the nearest neighboring pair of Fe ions.
Locations of Fe ions are represented by ${\bm r}_i = n_i^x{\bm e}_x + n_i^y{\bm e}_y +n_i^z{\bm e}_z$ for a set of integers $(n^{x}_{i}, n^{y}_{i}, n^{z}_{i})$ and unit vectors ${\bm e}_{x}$, ${\bm e}_{y}$, and ${\bm e}_{z}$ connecting adjacent Fe sites as shown in Fig. 1(a).
The non-zero terms of ${\bm p}_i$ in BiFeO$_3$ are originated from a $C_3$ point group \cite{Matsumoto2017} and any term of ${\bm p}^{\alpha}_{ij}$ is allowed to be non-zero from a symmetry argument \cite{Miyahara2016}. 
We defined the coupling constants of $p^\alpha_{ij}$ for one of the bonds along ${\bm e}_{z}$ as eqs. (S2) and (S3) in the Supplementary Material; ({\it e.g.}, $\Pi^X \equiv \Pi_{i i+z}^X$, $d^{XY} \equiv d^{XY}_{i i+z}$), and the other coupling constants for the inequivalent five bonds are expressed using the rotational-symmetry operation. 
\par The spin at the $i$-th site can be represented as ${\bm S}_i = \left( S\cos{\phi_{q}}\sin{{\bm q} \cdot {\bm r}_{i}}, S\sin{\phi_{q}}\sin{{\bm q}\cdot {\bm r}_{i}}, S\cos{{\bm q}\cdot {\bm r}_{i}} \right)$ and ${\bm S}_i = \{ S\sin{\left[\eta + (-1)^{n_{i}}\phi_{H}\right]}, -(-1)^{n_{i}}S\cos{\left[\eta + (-1)^{n_{i}}\phi_{H}\right]}, 0 \}$ in cycloidal and CAFM states, respectively, whereas $\bm{q} = \left( q_{0} \cos{\phi_{q}}, q_{0} \sin{\phi_{q}}, \sqrt{3} \pi / a_{\rm c} \right)$ and $q_{0} = 2\pi / \lambda$ with the cycloidal period $\lambda = 620$ {\AA} and lattice constant of the pseudocubic cell $a_{\rm c} = 3.96$ {\AA}. $\eta$ ($\ll 1$) represents the spin canting angle in the CAFM state.
In the cycloidal state, we ignore the minor spin component normal to the ${\bm q}$--${\bm Z}$ plane \cite{Ramazanoglu2011} for simplicity. 
\par By taking the spatial average in the cycloidal state, we can evaluate macroscopic polarizations as described in Supplementary Material.
In the cycloidal state, all mechanisms expressed in eqs. (1) and (2) can explain the emergence of spin-driven electric polarization ${\bm P}_{\rm mag}$ in the form 
\begin{align}\label{P1}
 &{\bm P}_{\rm mag} = \left( -P_{\rm t}^{0}\sin{2\phi_{H}}, -P_{\rm t}^{0}\cos{2\phi_{H}}, P_{Z} \right).
\end{align}
\begin{table}[t]
 \caption{Values of $P_{\rm t}^0$ in the eq. (\ref{P1}) of IDM, ES, and ML mechanisms in the cycloidal and CAFM phases, respectively. Details of these calculations are described in Supplementary Material.
 }\label{tab:para_table}
 \small
	\begin{tabular}{cccc}\\ \hline\hline
	  & IDM & EX & ML \\ \hline
	Cycloid & $\frac{\sqrt{6}}{4}q_{0}a_{\rm c}S^{2}\left( d^{XY} + d^{YX} \right)$ & $\frac{\left(q_{0}a_{\rm c} \right)^{2}}{2}S^2\Pi_{Y}$ & $-\frac{1}{2}S^{2}\Lambda^{X}_{XY}$ \\
	CAFM & $0$ & $0$ & $-S^{2}\Lambda^{X}_{XY}$ \\ \hline \hline
	\end{tabular}
\end{table}
Here, $P_{\rm t}^0$ in each phase is tabulated in Table I. 
As shown by the solid lines in Figs. 3(a)--3(c), eq. (\ref{P1}) reasonably reproduces the experimental results for the cycloidal state when setting $\phi_{H}=\phi_{q} + \pi/2$ and $P_{\rm t}^0 \sim 300 \,{\rm \mu C/m}^{2}$. 
In the CAFM phase, $\phi_{H}$ dependence is again expressed by eq. (\ref{P1}). Here, ES and IDM mechanisms suggest $P_{\rm t}^0 = 0$ contrary to the experimental results. 
On the other hand, the ML mechanism predicts doubling of the $P_{\rm t}^0$ in the CAFM phase. This predicted doubling of $P_{\rm t}^0$ together with the characteristic angular dependence in eq. (\ref{P1}) reasonably reproduce the experimental results as shown in Figs. 4(b) and 4(c). 
Therefore, spin-induced polarization in the cycloidal and CAFM phases can mainly be ascribed to the ML mechanism. 
\par In the ML mechanism, spin-driven electric polarization is determined by the local spin component at each Fe ion. 
For example, amplitude of $p_i^Y$ is proportional to $(S_i^X)^2$ when $\phi_H = \pi /2$ ($\phi_q = 0$) as shown in Supplementary Material. In the cycloidal state, spatial average of sinusoidally changing $(S_i^X)^2$, namely sinusoidally modulating $p_i^Y$, results in the reduction of $P_Y$ by a factor of 1/2 from that in the CAFM state. 
\par Finite magnetoelectric effects in the CAFM phase have been argued in some earlier reports \cite{Popov2004, Kadomtseva2004, Tokunaga2015JMMM, Kawachi2017}. 
Different from the earlier arguments, our results indicate emergence of the spontaneous electric polarization in the CAFM phase even in the zero-field limit, which is crucially important to utilize the CAFM states in thin films or Co-substituted BiFeO$_3$ at zero field \cite{Yamamoto2016, Sando2013}. 
Although we studied the magnetic control of ${\bm P}_{\rm t}$ in the CAFM phase at a temperature of 4.2 K, this phase exists up to a temperature higher than room temperature \cite{Tokunaga2010}. 
In addition, the $P_t^0$ remains finite in the limit of $\eta \rightarrow 0$ meaning simple collinear antiferromagnetic state can induce spin-driven polarization in this material. 
In many multiferroic materials, frustrated spin systems have been widely studied to realize complicated spin structures, which results in low ordering temperature. 
The present result suggests that magnetically controllable electric polarization of $\sim 600 \,{\rm \mu C/m}^{2}$ can be realized even in a simple antiferromagnet. This finding will pave a way to develop multiferroic materials useful at room temperature. 
\par In summary, we measured the angular dependence of field-induced changes in the electric polarization for a single crystal of BiFeO$_{3}$.
The electric polarization normal to the hexagonal axis was systematically controlled by changing the azimuthal angle of the applied magnetic field. 
Results reveal that this component is strengthened by the field-induced transition from the cycloidal phase to the canted antiferromagnetic phase. 
We explained this spin-driven component of electric polarization using the metal--ligand hybridization mechanism. 
\\
\par This work was supported by the JSPS Grant No. JP16K05413 and the MEXT Element Strategy Initiative to Form Core Research Center. 

\end{document}